# Shungite as loosely packed fractal nets of graphene-based quantum dots

Natalia N. Rozhkova[1] and Elena F. Sheka[2*]

[1]*Institute of Geology, Karelian Research Centre RAS, Petrozavodsk*

[2]*Peoples' Friendship University of Russia, Moscow*

E-mail: rozhkova@krc.karelia.ru

sheka@icp.ac.ru

**Abstract**

The current paper presents the first attempt to recreate the origin of shungite carbon at microscopic level basing on knowledge accumulated by graphene science. The main idea of our approach is that different efficacy of chemical transformation of primarily generated graphene flakes (aromatic lamellae in geological literature), subjected to oxidation/reduction reactions in aqueous environment, lays the foundation of the difference in graphite and shungite derivation under natural conditions. Low-efficient reactions in the case of graphite do not prevent from the formation of large graphite layers while high-efficient oxidation/reduction transforms the initial graphene flakes into those of reduced graphene oxide of ~1nm in size. Multistage aggregation of these basic structural units, attributed to graphene quantum dots of shungite, leads to the fractal structure of shungite carbon thus exhibiting it as a new allotrope of natural carbon. The suggested microscopic view has finds a reliable confirmation when analyzing the available empirical data.

## 1 Introduction

Carbon is an undisputed favorite of Nature that has been working on it for billions years thus creating a number of natural carbon allotropes. Representatively, we know nowadays diamond, graphite, amorphous carbon (coal and soot), and lonsdaleite. To some extent, fullerene $C_{60}$ of the extraterrestrial origin can be attributed to this group as well. Recently revealed growth mechanism by incorporation of atomic carbon and $C_2$ has shed new light on the fundamental processes that govern self-assembly of carbon networks, thus illuminating astrophysical processes near carbon stars or supernovae that result in $C_{60}$ formation throughout the Universe[1]. For the last three decades, the allotrope list has been expanded over artificially made species such as fullerenes, single-walled and multi-walled carbon nanotubes, glassy carbon, linear acetylenic carbon, and carbon foam. The list should be completed by nanodiamonds and nanographites, to which belong one-layer and multi-layer graphene. Evidently, one-to-few carbon layers adsorbed on different surfaces should be attributed to this cohort as well.

Looking at the whole carbon family, one can notice that practically each of the natural allotropes not only possesses own set of characteristics but is duplicated in an artificially made counterpart. Thus, diamond and graphite present cubic and hexagonal crystalline modifications governed by $sp^3$ and $sp^2$ electronic configurations of carbon atoms, respectively while nanodiamonds and graphenes keeping the same configurations offer particular size effects. Lonsdaleite is a $sp^3$ configured hexagonal allotrope of the carbon, believed to form when meteoric graphite falls to the Earth. One-to-few atomic adsorbed carbon layers with regular hexagonal structure can be obtained on different surfaces, such as Si(111), Ag(111), and so forth, thus presenting a nanosize counterpart of this natural carbon. Coal and soot do not have any crystalline structure and both are informally called amorphous carbons. They are products of pyrolysis, which does not produce true amorphous carbon under normal conditions, and have a complicated substructure. Fullerenes, carbon nanotubes, carbon clusters, and linear acetylenic carbon (graphynes and graphdynes) fit nicely the substructure elements.

In spite of high abundance of the carbon allotropes, the above list remains incomplete until shungite is added to the group of natural allotropes. As has been known, this natural carbon deposit cannot be attributed to either diamond, or graphite and amorphous carbon. A lot of efforts have been undertaken to exhibit that the material, once pure carbon by content, presents a fractal structure of agglomerates consisting of nanodimensional globules[2], each of which presents a cluster of ~ 1 *nm* in size graphene-like flakes[3].

The current paper presents a summarized view on shungite as a new carbon allotrope and suggests a microscopic vision of the shungite derivation supplemented by the presentation of its structure as a multistage fractal nets of reduced graphene oxide (rGO) flakes.

**2 What we know about shungite**

Shungite rocks are widely known and are in a large consumer's demand due to its unique physico-chemical[4] and biomedical properties[5]. For a long period, thorough and systematic studies, aimed at clarification of the reasons of such uniqueness, have been performing. As has been gained[2], shungite carbon of natural deposits is a densely packed porous structure of agglomerates with a large variety of pore size from units to hundreds *nm*. Such a large dispersion of the pore size evidences a multistage structure of agglomerates that points to the fractal accommodation[6]. The shungite fractal structure has been clearly evidenced by Small Angle Neutron Scattering (SANS)[7]. SANS showed as well that there were two types of shungite pores, namely, small pores with linear dimensions of 2-10 *nm* and large pores of more than 100 *nm* in size. Taking together, the findings have allowed suggesting that the shungite fractal structure is provided with aggregates of globular particles of <6 *nm* in average size. In its turn, the globules are clusters of graphene-based fragments of ≤1 *nm* in size[8,9].

A high ability of shungite to be dispersed up to individual aggregates in aqueous solution convincingly proves this structure vision. Such dispersions, with maximum achieved concentration of ~0.1mg/ml, are described in details elsewhere[2,8]. If the water evaporation is blocked, the dispersions are quite stable and conserve properties during a long period of time (for several years) (see details in Electronic Supplement Information, ESI). A complete drying of the dispersions results in the formation of densely packed powdered condensate. The structural and physico-chemical characteristics of the latter are quite identical to those of the pristine shungite[8]. In its turn, Raman scattering of the dispersions (see Fig. 2S) convincingly evidences the graphene-based origin of basic structural units of the shungite aggregates.

The aqueous dispersions exhibit a large number of peculiar properties that, on one hand, have a direct connection with the unique properties of shungite while on the other, are pretty similar to those characteristic for aqueous dispersions of such quantum dots as either Ag and Au nanoparticles or CdS and CdSe nanocompositions, on one hand, and synthetic graphene quantum dots (see exhausted review[10]), on the other. Similarly to the former, shungite dispersions reveal high activity towards enhancing nonlinear[11-13] and spectral[14] optical properties. Analogously to the latter, shungite dispersions exhibit a close similarity in the appearance of a high inhomogeneity of both morphological and spectral properties. A particular attention should be given to their biomedical behaviour[5,15,16]. Thus, the study of the dispersion effect on the behaviour of serum albumin has shown that the shungite globules and proteins form stable bioconjugates. The latter do not change the protein secondary structure, but causes a drastic lowering of the compactness of the protein tertiary structure that might promote various biomedical applications. All of these properties have allowed for speaking about shungite dispersions as colloidal systems of 'graphene quantum dots of shungite' (GQD-SH) related to 'graphene quantum dots of natural origin'[14] and about shungite itself as a new carbon allotrope consisting of loosely packed fractal nets of GQD-SH. Therewith, shungite presents the natural form of the allotrope while the aqueous-dispersion condensate is one of the post-treated products.

**3 Graphene molecular chemistry defines the origin of shungite carbon**

The graphene-based basic structure of shungite provides a good reason to consider the latter at the microscopic level by using high power of the modern empirical and theoretical molecular science of graphene. This approach allows us not only to explain all the peculiarities of the shungite behavior, but also to lift the veil on the mystery of its origin. To consider shungite from the viewpoint of molecular science of graphene, in fact, is to find answers to the following questions.

1. Where did graphene-based basic units of shungite come from?
2. Why is the unit linear size limited to ~1 *nm*?
3. What did this size stabilize during the geological time of life?
4. What is the chemical composition of the basic units?
5. Why and how do the units aggregate?
6. Why there are two sets of pores in the shungite carbon?
7. What is meant by graphene quantum dots of shungite?

More recently, each of the problems mentioned above has been the topic of a separate study and attempts to consider them all together have seemed completely unrealistic. However, today the situation has changed drastically. Knowledge that the molecular graphene science has accumulated for the last few years is so vast that it allows considering the totality of issues simultaneously. Obviously, not all the answers to the above questions have been so far fully exhaustive. However, they present the first attempt of seeing the problem as a whole leaving details of the subsequent refinements for future investigations. Currently, the following answers can be suggested.

**Answer 1**. To answer the first question, we have to address the geological story of shungite. Although shungite is about two billions years old, its origin has been still under discussion[17]. The available hypothesizes are quite controversial. According to the biogenic concept, it is formed of organic-carbon-rich sediments. Following the others, shungite is of either volcanic endogenous or even extraterrestrial origin. In contrast to graphite, which is largely distributed over the Earth, the shungite deposits are space limited, and the Onega Lake (OL) basin of Karelia is the main site for the rock mining.

Two distinct peculiarities are characteristic for the geological Karelian region, namely: 1) shungite deposits around the OL neighbor with graphites in the vicinity of the Ladoga Lake (LL); and 2) the abundance of water, both open and mineral one, in the former case. The first feature gives clear evidence that the Karelian region as a whole is favorable for graphene-like deposition of carbon which might imply the presence of a common framework of the two geological processes. The second forces to draw a particular attention to the aquatic environment of the deposition.

Geology of graphite is well developed by now. According to the modern concept[18], graphite can be (i) either syngenetic, formed through the metamorphic evolution of carbonaceous matter dispersed in the sediments or (ii) epigenetic, originating from precipitation of solid carbon from carbon-saturated C–O–H fluids. The privilege is given to the first one. The transformation of carbonaceous matter involves structural and compositional changes of basic structural units in graphite in the form of aromatic lamellae (graphene flakes) and occurs in the nature in the framework of thermal or regional metamorphism that apart from temperature, involves shear strain and strain energy. Pressure and shear promote molecular ordering of the lamellae and facilitate preferential alignment of the latter as well as pores coalescence. (A high efficacy of graphene to re-knit its holes has been recently shown under laboratory conditions[19]). Temperature and pressure efficiently govern the graphitization. According to Ref. 20 and 21, graphitization begins and ends at temperatures from 380 to ~450 $^0$C and pressures between $2 \times 10^8$ Pa (2 kbar) and $3 \times 10^8$ Pa (3 kbar).

Accepting the syngenetic graphitization to be a common process for the derivation of both graphite and shungite in the Karelian region, we can suggest the following answer to the first question.

The graphitization is a longtime complicated process that, occurring during a geological scale of time, can be subjected to various chemical reactions. Tempo and character of the reactions are obviously dictated by the environment. It is quite reasonable to suggest that the aqueous environment at 300-400$^0$C, under which the metamorphic evolution of carbonaceous matter occurs, is a dynamically changeable mixture of water molecules, hydrogen and oxygen atoms, hydroxyl and carboxyl radicals as well. The interaction of the carbonaceous matter, subjected to structural and compositional changes in due course of alignment of graphene lamellae and pores coalescence with this mixture accompanies the process. The most expected reactions concern hydrogenation, oxidation, hydration, hydroxylation, and carboxylation of the formed lamellae. At this point, it is important to note that, according to the molecular theory of graphene[22-27], any reaction of these kinds primarily involves edge carbon atoms of the flakes. Actually, Fig. 1 presents a typical image map of the atomic chemical susceptibility (ACS) distribution over a graphene flake atoms (the flake is presented by a rectangular graphene fragment, below (5, 5) NGr molecule, containing $n_a$=5 and $n_z$=5 benzenoid units along armchair and zigzag edges, respectively). The ACS map shape is characteristic for a graphene fragment with bare edge atoms of any size and shape. Presented in Fig.1 shows that the chemical reactivity of the edge atoms exceeds that one of the basal plane atoms ~4 times so that any addition reaction will start by involving these atoms first. The addition, obviously, terminates the flake growth thus preventing the large-layer graphite formation. Empirically, it has been repeatedly observed in the case of graphene oxide (GO)[28-31]. Therefore, since the above mentioned reactions start simultaneously with the deposit formation, their efficiency determines if either the formed graphene lamellae will increase in size (low efficient reactions) or the lamellae size will be terminated (reactions of high efficiency). Since large graphite deposits are widely distributed over the Earth, it might be accepted that aqueous environment of the organic-carbon-rich sediments generally does not provide suitable conditions for the effective termination of graphene flakes in the course of the deposit graphitization. Obviously, particular reasons may change the situation that can be achieved in some places of the Earth. Apparently, this occurred in the OL basin, which caused the replacement of the graphite derivation by the shungite formation. Some geologists reported on correlation of the formation of shungite with the increase of oxygen concentration in atmosphere that occurred in 1.9-2.1 Ga[32]. This fits well the geochemical boundary of the Earth history of 2.2 Ga.

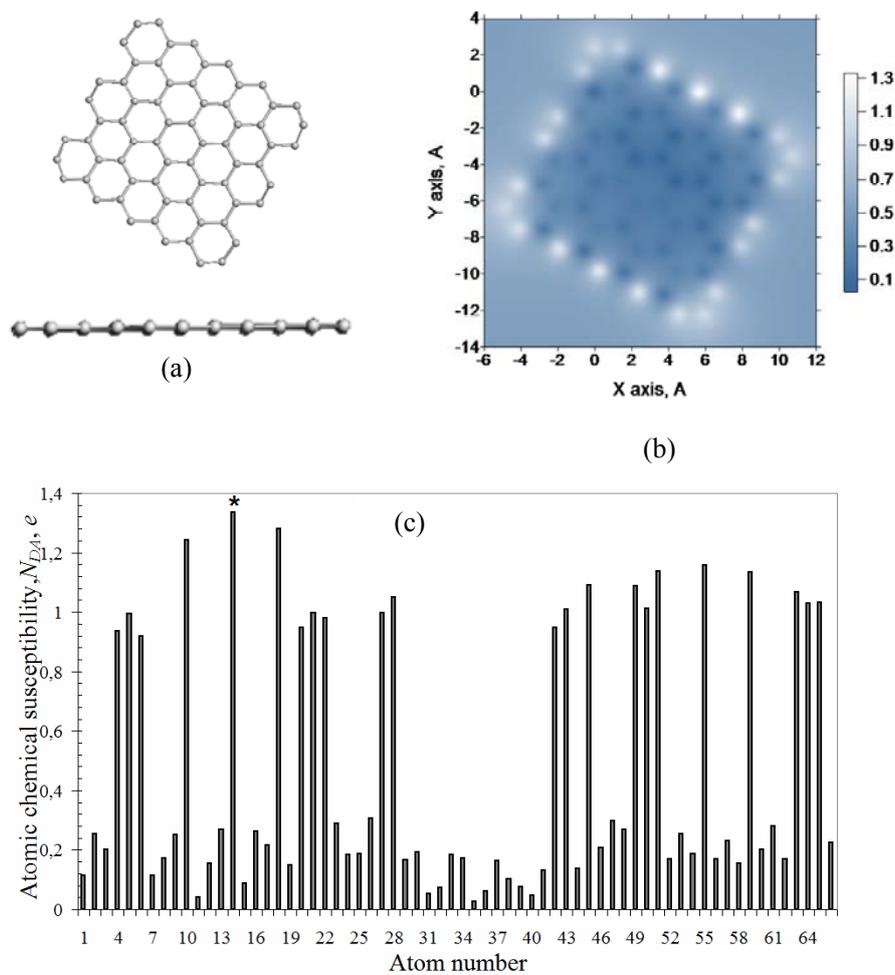

**Figure 1**. Top and side views of the equilibrium structure of the (5,5) NGr molecule (a) and atomic chemical susceptibility distribution over atoms in real space (b) and according to atom numbers in the output file (c) (UHF calculations[25]).

**Answer 2**. If chemical modification of graphene lamellae is responsible for limiting their size, the answer to the question about the size limitation to ~1 *nm* should be sought in the relevant reaction peculiarities. First of all, one must choose among the reactions that are preferable under the above graphitization conditions. Including hydroxylation and carboxylation into oxidation reaction, we must make choice among three of them, namely, hydration, hydrogenation, and oxidation. All the three reactions are well studied for graphene at molecular level both empirically and theoretically.

The pristine graphene is hydrophobic so that its interaction with water molecules is weak. Chemical coupling of water molecule with graphene can rarely occurs at the zigzag edge and is characterized by small coupling energy (see ESI). According to this, water cannot be considered as a serious chemical reactant responsible for the chemical modification of the pristine graphene lamellae. Nevertheless, water plays extremely important role in the shungite fortune that will be discussed later.

At molecular level, depending on external conditions concerning the fixation of the flake perimeter and the accessibility of the flake basal plane to hydrogen atoms either from one- or two-sides, different graphene hydrides (GHs) are formed[25] (see Fig. 3S of the ESI). It should be pointed that each GH is a polyderivative substance. It was shown that graphene hydrogenation is a quite active chemical reaction in the course of which not molecular but atomic hydrogen can be chemically bonded with both edge and basal plane atoms. However, empirically, as shown by active studying of the reaction (see[33] and references therein), the graphene hydrogenation is a difficult task, and the process usually involves such severe conditions as high temperatures and high pressure or employs special devices, plasma ignition, electron irradiation, and so forth. One of the explanations can be connected with the necessity in overcoming a barrier at each addition of the hydrogen atom to graphene body. Figure 4S demonstrates the dependence of coupling energies of different addends on these addends distance from the targeted carbon atom at the zigzag edge of the (5, 5) NGr molecule. In the case of hydrogen, the plotting clearly reveals the barrier that constitutes ~13 kcal/mol. Therefore, graphene hydrogenation can hardly play the main role during shungite formation under conditions mentioned earlier.

In contrast, graphene oxidation can apply for a role. The reaction is studied thoroughly at different conditions (see reviews[28-31] and references therein) and the achieved level of its understanding is very high. The latter has led the foundation of massive fabrication of a particular 'graphene' that is reduced graphene oxide (rGO). The oxidation may occur under conditions that provide the shungite derivation in spite of low acidity of the aqueous surrounding but due to long geological time and practically barrierless character of the reaction concerning additions of either oxygen atoms or hydroxyls to the graphene body as seen in Fig. 4S. As shown, oxidation causes a destruction of the pristine graphite and graphene sheets just cutting them into small pieces[29, 34]. Thus, 900 sec of continuous oxidation cut a large graphene sheet into pieces of ~1 nm in size[29]. Important, that further prolongation of the oxidation does not cause decreasing the size thus stabilizing them at the 1 nm level. This finding allows suggesting that shungite flakes of ~1 nm in size have been formed in due course of geologically prolong oxidation of graphene lamellae derived from the graphitization of carbon sediments.

**Answer 3**. Numerous experimental studies (see[29-31, 35-37]) and a recent detailed consideration of graphene oxides (GOs) from the viewpoint of molecular theory[26] have shown that GOs are products of the hetero-oxidant reaction. Three oxidants, among which there are oxygen atoms O, hydroxyls OH, and carboxyls COOH, are the main participants of the process albeit in different ways by participating in the formation of the final product. In Fig.2a is presented the final product of the (5, 5) NGr molecule oxidation that follows from the molecular theory of graphene. The structure has been obtained in the course of the stepwise addition of the above oxidants to the molecule under the conditions that the molecule basal plane is accessible for the oxidants from the top only. The choice of the preferable oxidant was made following the criterion of the largest per step coupling energy[26]. Carboxyl units, located both at the edge of the molecule and at its basal plane, do not meet the criterion and lose a competition to two other oxidants. Once currently investigated, increasing the molecule size allows for revealing a small fraction of the units in the molecule edge area only (see Fig. 6S). Plottings in Fig.2b correspond to the per step coupling energies that describe the energetics of the oxidants attachment.

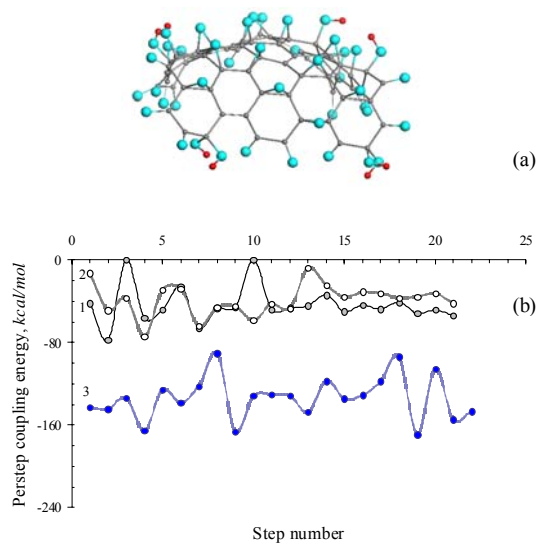

**Figure 2.** Based on the (5,5) NGr molecule, structural model of a top-down exfoliated GO ((5,5) GO molecule) (a) and per step coupling energy (b) versus step number for this GO family under subsequent O- and OH- additions to carbon atoms at either the molecule basal plane (curves 1 and 2) or edges (curve 3)[26]. Dark gray, blue, and red balls mark carbon, oxygen, and hydrogen atoms, respectively.

Hydrothermal conditions of the shungite derivation present serious arguments in favor of a hypothesis about GO origin of the deposits. However, the hypothesis strongly contradicts the atomic percentage of oxygen in the carbon-richest shungite rocks that constitutes ~ 1.0 %[38] instead of

30-45% expected from the GO concept. This contradiction forces to think about of a complete or partial reduction of the preliminary formed GO occurred during geological process.

As follows from the plottings in Fig. 2b, GO is characterized by two regions of chemical bonding of oxidants with graphene body. While the edge atoms region should be attributed to that one of strong chemical bonding, the basal plane, for which the average coupling energy is three times less, is evidently related to the area of a weak coupling. The finding is crucial for the GO reduction showing that the latter concerns basal plane in the first instance while oxidants located in the framing area may not be removed under conditions of the convenient reduction without destruction of the carbon skeleton. This finding explains the residual oxygen content in reduced rGOs of 5-10%[29-31, 39]. As seen, shungite is no exception to this pattern and its ~1.0% content of oxygen fits into the overall picture related to rGOs.

Usually a synthetic GO reduction occurred when using chemically strong reductants that are not available in the natural environment of shungite. However, as has been recently shown, the rGO can be obtained just in water, which only requires a much longer time for the reduction[40]. Evidently, the geological time of the shungite derivation might be quite enough for the reduction of pristine GOs in water.

**Answer 4.** Figure 3 presents the equilibrium structure of the (5, 5) rGO molecule that has 1.3 x 1.4 $nm^2$ cross section. The molecule was obtained in the course of the structure optimization after removing all epoxy and hydroxy groups from the basal plane of the (5, 5) GO shown in Fig.2[26]. Due to recovering $sp^2$ configuration for carbon atoms at basal plane, the rGO molecule noticeably regenerates its planarity, although impaired, especially in the corner areas. Basing on empirical estimation of ~1 $nm$ for a basic shungite graphene-based flake, the (5, 5) rGO molecule might be considered as one of possible configurations of the basic shungite structural unit. However, the atomic percentage of oxygen in the case constitutes 33% that is far from the empirical contamination. The controversy may mean that the actual oxygen framing of rGO flakes is not fully saturated. A large oscillation amplitude of the per step coupling energy related to the framing area (see curve 3 in Fig. 2b) may be one of possible reasons. Actually, the atoms, which correspond to the top part of the plotting, may be removed during the reduction additionally to the basal ones. Another reason can be connected with the stability of the rGO flakes that depends on the flakes shape and corner structure, on one hand, and thermodynamic conditions of the oxygen reservoir, on the other[41]. At any rates, the problem needs a further examination.

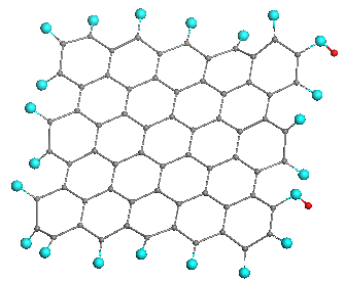

**Figure 3.** Based on the (5,5) NGr molecule, structural model of a reduced top-down exfoliated GO[26] ((5, 5) rGO molecule).

**Answer 5**. Assuming that nanosize rGO flakes generated in aqueous media present the first stage of shungite derivation, let us trace their path from individual molecules to densely packed shungite carbon. Obviously, the path is through successive stages of the flakes aggregation. Empirically was proven that aggregation is characteristic for both synthetic GO and rGO flakes. Thus, infrared absorption[42] and Inelastic Incoherent Neutron Scattering (IINS)[43, 44] have shown that synthetic GO forms stacked structures that confine water. Just recently, a similar picture has been obtained for synthetic rGO[44]. Neutron diffraction has shown therewith[44] that the characteristic graphite interplanar distance $d_{002}$ constitutes, in average, ~6.9 Å and ~3.5Å in the case of GO and rGO, respectively, evidently indicating the recovery of the GO carbon carcass planarity due to its reduction. Computationally, it was confirmed in the current study that water molecules are comfortably located between the neighboring layers of the GO stacked structure while none of the water molecules can be retained near the rGO basal plane (see Fig. 4). Regardless of the molecule starting position, each of them is displayed outside the rGO basal plane area once kept in the vicinity of framing atoms. The finding is well correlated with the rGO short-packed stacked structure leaving the place for water molecules confinement in pores formed by stacks.

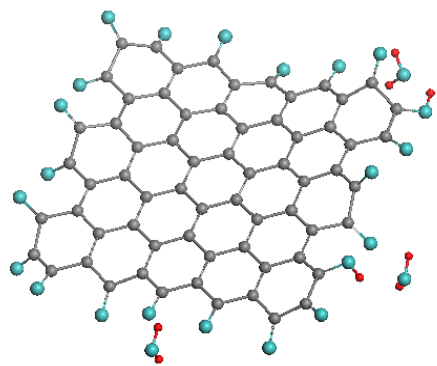

**Figure 4.** Equilibrium structure of complex involving (5,5) rGO molecule and three water molecules. Initially, water molecules were accommodated within the framework of the basal plane. The total coupling energy constitutes -27.15 kcal/mol.

Neutron diffraction and IINS study of both pristine shungite and its dense condensate obtained in the course of evaporation of shungite water dispersions have shown a close similarity in the behavior of shungite and synthetic rGO[45]. The average $d_{002}$ value of both shungite powders constitutes ~3.5Å while IINS evidences the presence of confined water, whose quantity, however, considerably exceeds that one of the synthetic rGO. The finding firstly confirms that the rGO flakes present the basic structural units of shungite and secondly allows for suggesting a vision of the next stages of the shungite structure towards shungite rocks.

The characteristic $d_{002}$ diffraction peaks of shungite are considerably broadened in comparison with those of graphite which allows for estimating approximate size of the rGO flakes stacks of ~1.5 $nm$[45]. The stacks form the second-stage structure of shungite. Obviously, their irregular distribution in space causes the formation of different pores. Some models of the pores are shown in Fig.5. The inner surface of the pores is carpeted with oxygen atoms that can willingly hold water molecules in their vicinity as follows form Fig. 4. Thus retained water is the main consent of the registered IINS spectra[45]. As seen in Fig.5, the pore formation results in structurally distinguished combinations of stacks around the pores. These combinations present the third stage of the shungite fractal structure and are attributed to globules[2] of a few $nm$ in size. Further aggregation of the globules leads to the formation of the bigger aggregates with linear dimensions of 20-100 $nm$ as follows from their size distribution shown in Fig. 1S. The aggregate agglomeration completes the formation of fractal structure of shungite.

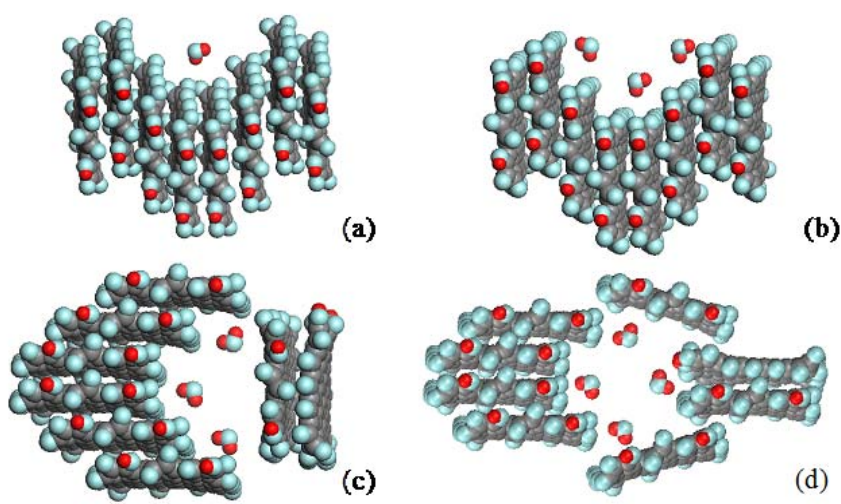

**Figure 5.** Sketches of primary shungite pores and their fragments on the basis of the (5,5) rGO molecule. *a* and *b*. Stack of 8 rGO molecules forms a concave cavity capable to hold a few water molecules. The average interplanar distance $d_{002}$ in the stack and the stack length constitute 3.85Å and ~15Å, respectively. *c* and *d*. Possible configurations of the pores inner structure
.

**Answer 6**. In fractal structures that are rich in pores, the pores size is usually tightly connected with the size of structural elements involved in the pore formation[46]. Moreover, the larger is variety of the elements size and structure, the bigger is distribution of the pores over size. In view of this, the different size of the multistage shungite fractal structural elements evidently predetermines different sizes of shungite pores. Thus, as seen in Fig.5, one of linear size of the pores formed by rGO flakes is determined by linear dimensions of the flakes while two others are defined by the thickness of the flakes stacks. Therefore, basic rGO flakes and their stacks are responsible for shungite pores of 2-5 $nm$ in size. Following this line, globules may form pores up to 10 $nm$ while extended aggregates of globules obviously form pores of a few tens $nm$ and bigger. This presentation well correlates with the SANS experimental data that evidence the presence of two sets of pores in shungite in the range of 2-10 $nm$ and above 100 $nm$[7]. In its turn, a detailed analysis of the deviation of the water IINS spectra of shungite from that one related to ice points to the water confining in pores of 2-3 $nm$ in size[45].

Taking together, the multitier of structural elements and various porosity make the fractal structure of shungite good self-consistent.

**Answer 7**. Answering questions from 1 to 6, we have been passing all stages of the shungite derivation starting from naked graphene lamellae, going through their oxidation and reduction, and completing the process by aggregation resulted in the formation of the shungite carbon fractal structure of a multilevel hierarchy. It would seem that the latter circumstance makes it difficult to determine the quantum dot of shungite. However, it should be recalled that the feature of fractal structures is that fractals are typically self-similar patterns, where "self-similar" means that they are "the same from near as from far"[46]. This means that the peculiarities of, say, optical behaviour of all the multistage structural units obey a single law. From this viewpoint, there is no difference which structural element of shungite should be attributed to a quantum dot. Obviously, the basic structural units have certain advantages among other multistage elements that is why graphene quantum dots of shungite (GQD-Sh) should appear to associate with rGO individual flakes. A similar problem takes place concerning synthetic graphene quantum dots[10]. In fact, a solute of GQDs in aqueous solutions represent a sparse fractal structure, basic structural elements of which are individual rGO flakes. Therefore, synthetic GQDs and GQDs-Sh are of the same origin. The latter has been recently confirmed when observing a close similarity of photoluminescence spectra of aqueous solutions of synthetic GQDs and aqueous dispersions of shungite[14]. This finding has added one more argument in favor of the rGO origin of GQD-Sh.

**4 Conclusive remarks**

When the current paper has been written, the authors became familiar with a fascinating overview on "Small but strong lessons from chemistry for nanoscience", written by one of the most famous chemist of our time, Roald Hoffmann[47]. Hoffmann's lessons are in tight connection with the problem discussed above, just more strictly highlighting the sharpest points of the topic and giving a strong support to the approach suggested by the authors.

The main idea of our approach is that molecular chemistry lays the foundation of the difference in graphite and shungite derivation under natural conditions in due course of the geological time. The idea showed the way of checking this suggestion by exhibiting chemical reactions that might be responsible for the deposits derivation as well as by simulating final products of the relevant reactions. Based on the theory of graphite genesis18, possible reaction components involve molecular objects simulating fragments of polycondensed carbon molecules (carbon substrate, or naked graphene lamellae), on one hand, and molecular (water, carboxyl, hydroxyl) and atomic (hydrogen, oxygen) species (chemical reactants), on the other. The naked graphene lamellae are kinetically unstable, as Roald Hoffmann called such objects[47], since covalent bonds are cut at their edges. We described this instability in terms of atomic chemical susceptibility $N_{DA}$ that points to local radicalization[22-27]. Such species, according to[47], 'will try to heal themselves' and external molecules may stabilized them. In full agreement with our suggestion, Hoffmann continues that "too great stabilization will inhibit growth; too little stabilization will not prevent from collapse to the solid". The difference in the graphene lamellae stabilization was the second basic idea of our approach. Thus, the great stabilization of the lamellae results in the shungite formation while the little one provides derivation of graphite deposits. The stabilizing reactions are controlled by the reactants coming on and off the pristine naked graphene lamellae. Fully agreeing with the opinion of Hoffmann, we believe that both thermodynamic (Gibbs energies) and kinetic (activation energies) factors matter in the dynamic process. Shungite is suggested to be formed due to balance of a number of multi-reactant processes, each governed by its own thermodynamics and kinetics. The presence of other elements such as silicon and metals undoubtedly influences the deposit formation. Actually, the Karelian deposits of shungite are non uniform by the carbon content, value of which changes from 3% to 98 wt%[2]. Silicon is the main partner of the mixed depositions. However, speaking about new allotrope we imply a particular shungite rock from the Shun'ga deposit with the highest carbon content up to 98.0%[38] for which the presence of other Earth element is negligible[48].

These two main concepts have been considered in the paper addressing oxidation/reduction reactions that govern chemical modification of the pristine graphene lamellae. Basing on a wide experience gained for graphene chemistry in the laboratory conditions and an extended computational system experiment performed earlier[26], the oxidation/reduction reactions are shown to have a big privilege against hydration and hydrogenation of graphene. The two reactions work simultaneously but serving different purposes: oxidation stabilizes the growth of graphene lamellae thus determining their size, while reduction releases the oxygenated flake from weakly bound reactants located through over basal plane of the lamella leaving other located at the lamella circumference thus preserving the lamella stabilization. This conclusion is in full agreement with shungite empirical data related to exhibiting 1) ~1*nm* planar-like flakes of reduced graphene oxide as the basic structural element of the macroscopic shungite structure; and 2) remaining ~1 at% content of oxygen in the most carbon-pure shungite samples.

Shungite is formed in aqueous surrounding and although water molecules do not act as active chemical reactants, they play a very important role in composing shungite as a solid. Firstly, water medium acts as the reductant[40]. The slow rate of reduction evidently favors the accumulation of rGO flakes during a long shungite geological story. Secondly, water molecules fill the pores, helping to strengthen the framework of fractal shungite carbon.

Represented attempt to recreate the origin of the shungite at microscopic level was made possible by large knowledge, both empirical and computational, that has been accumulated by the graphene science. At the same time, its successful realization has shown that the processes occurring in Nature in a macroscopic scale are subordinated to the same laws as observed in a nanoworld and Chemistry which is the atomic-level science has ruled score. In this connection, it is worthwhile to finish the paper by conclusive words from the Hoffmann overview[47]: "What we have learned in chemistry, beautiful knowledge, gained without waiting for microscope to let us physically see down in there, of course applies to nano-objects. It's one world".

**Acknowledgement:** The work was supported by Basic Research Program, RAS, Earth Sciences Section-5 and grant RFBI 13-03-00422.

**Shungite as loosely packed fractal nets of graphene-base quantum dots**

Natalia N. Rozhkova[1], Elena F. Sheka[2]

[1] Institute of Geology, Karelian Research Centre, Russian Academy of Sciences, Petrozavodsk, Russia
[2] Peoples' Friendship University of Russia, Moscow, Russia

**Shungite dispersions preparation**. Stable aqueous dispersions of shungite with carbon concentration of 0.1 mg/ml are produced by ultrasonic treatment of the shungite powder (98 wt.% carbon) followed by filtration and ultracentrifugation in accordance with the technological protocol described in[1]. The size distribution (SD) of colloidal particles is controlled by the dynamic light scattering (DLS) using Zetasizer Nano ZS (Malvern Instruments) nanoparticle size analyzer. The size distribution pattern depends on the centrifuging regimes. However, the general view of SD profiles is rather stable and fits well the spectra of round particles. One of the typical SD profiles of the shungite aqueous dispersion is presented in Figure 1Sa. The profiles usually are wide with FWHM of 25.5 nm in the given case. The average colloidal particles size is 53.6 nm. Big ratio of the profile width to its maximum is a typical for highly inhomogeneous dispersions with respect to the size and possible configuration of the colloidal particles. Characteristics of each individual dispersion remain unchanged for more than one year, unless the solvent evaporates.

Condensation of the colloidal particles of aqueous dispersions at ambient conditions provides the formation of pronounced network structures. A complete drying of the dispersions has resulted in producing a densely packed powdered condensate (Fig.1Sb and c). Its structural and physico-chemical characteristics are quite identical to those of the natural shungite carbon[2].

**Raman scattering testing.** Figure 2S shows Raman spectra of two aqueous dispersions as well as of dispersion condensate recorded on a Nicolet Almega XR (Thermo Scientific) spectrometer at laser excitation of 532 nm. The instrument spectral resolution is 1 cm$^{-1}$ and the laser power is 300 mW. Valent O-H oscillations of ~3400 cm$^{-1}$ dominate in scattering and Fig. 2S presents only its low-frequency part whose recording requires 30-minute accumulation of the data. As seen in the figure, two well known graphene-characteristic G and D bands at 1594.6 cm$^{-1}$ and 1339.7 cm$^{-1}$ present the main features of the condensate spectrum indicating graphene-related structure of its main structural elements. A big $I_D/I_G$ ratio evidences a considerable difference of these elements structure from the ideal graphene[3]. The spectrum coincides with that of initial shungite powder. Analogous two bands appear to dominate in spectra of both dispersions. However, if bands at 1339.7 cm$^{-1}$ strictly coincide with D band of the condensate and can be attributed to similar bands characteristic for shungite colloidal particles, the S bands seemingly analogues to G band but located at 1633 cm$^{-1}$ in the dispersion spectra are too wide, both largely shifted to high frequencies, and too intense, which causes inverting the $I_D/I_G$ ratio. Since, evidently, graphene-based fragments of the dispersions are not more regular than in the condensate, the shift, intensity and width of these bands should be attributed to Raman scattering from other non-graphene vibrational excitations. It is quite reasonable to suggest the scissor deformational vibrations of water play the role. A shift from the frequency of 1585 cm$^{-1}$ related to free water molecules to 1633 cm$^{-1}$ in the dispersions points to hydrated state of water[4]. The latter is confirmed by the appearance of these vibrations in the observed Raman spectra, once symmetrically forbidden for free water. Basing on the intensity of D bands in the dispersion spectra, it is evident that only a small part of S bands is provided by scattering from shungite colloidal particles.

**Computational technique**. All computations have been performed in the broken symmetry approach by using unrestricted Hartree-Fock computational scheme implemented in CLUSTER-Z1 codes based on semiempirical AM1 approach (a detailed description of the strategy of the computational consideration of $sp^2$ nanocarbons is summarized in[5]).

**Graphene hydrogenation**. At molecular level, depending on external conditions concerning the fixation of a graphene flake circumference and the accessibility of the flake basal plane to hydrogen atoms either from one- or two-sides, different graphene hydrides (GHs) are formed[6]. Figure 3S presents a collection of graphene hydrides obtained under different conditions. It should be pointed that each GH is a polyderivative substance. Empirically, GHs shown in panels *a* and *c* have been observed for graphene membranes located over $SiO_2$ substrate[7].

**Barriers determination**. Figure 4S presents equilibrium structures of the H-, O-, OH-, and COOH_monoderivatives of the (5, 5) NGr molecule. The calculated barrier profiles, related to the desorption of the addends from the molecule, are plotted in Figure 5S. The calculations were performed at constant 0.05Å elongation of each relevant bonds.

**Water adsorption**. Figure 6S presents equilibrium structures of the (5, 5) NGr molecule interacting with one water molecule. As seen in the figure, water molecule is weakly coupled with the graphene molecule in all the cases supporting a well known physical nature of the adsorption events. The coupling energy varies from -4. 64 to +6.81 kcal/mol. The strongest coupling is observed for water attached to zigzag edge. When initial C-O distance is less than 1.8E, the water molecule form a weak chemical bond (C-O bond is of 1.50Å in length) with graphene. However, the coupling energy excels that one when water molecule is far away from graphene (Fig. 6Sa) at 0.84 kcal/mol only.

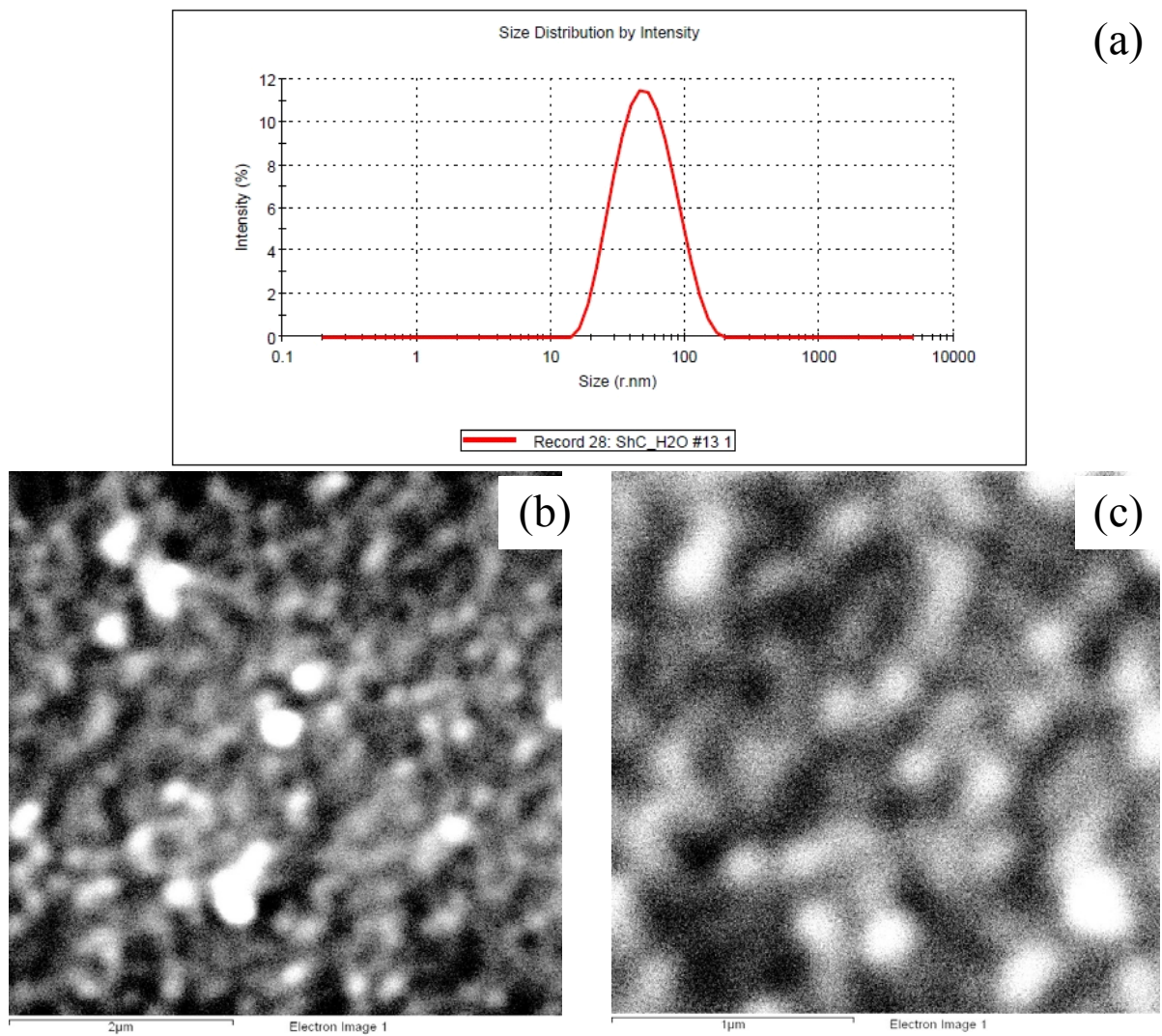

**Figure 1S.** Size distribution of colloidal particles in shungite water dispersion (a) and SEM images of shungite condensate on glass substrate (b) and (c): Scale bar 2 μm and 1 μm, respectively. Carbon concentrations constitute 0.1 mg/ml.

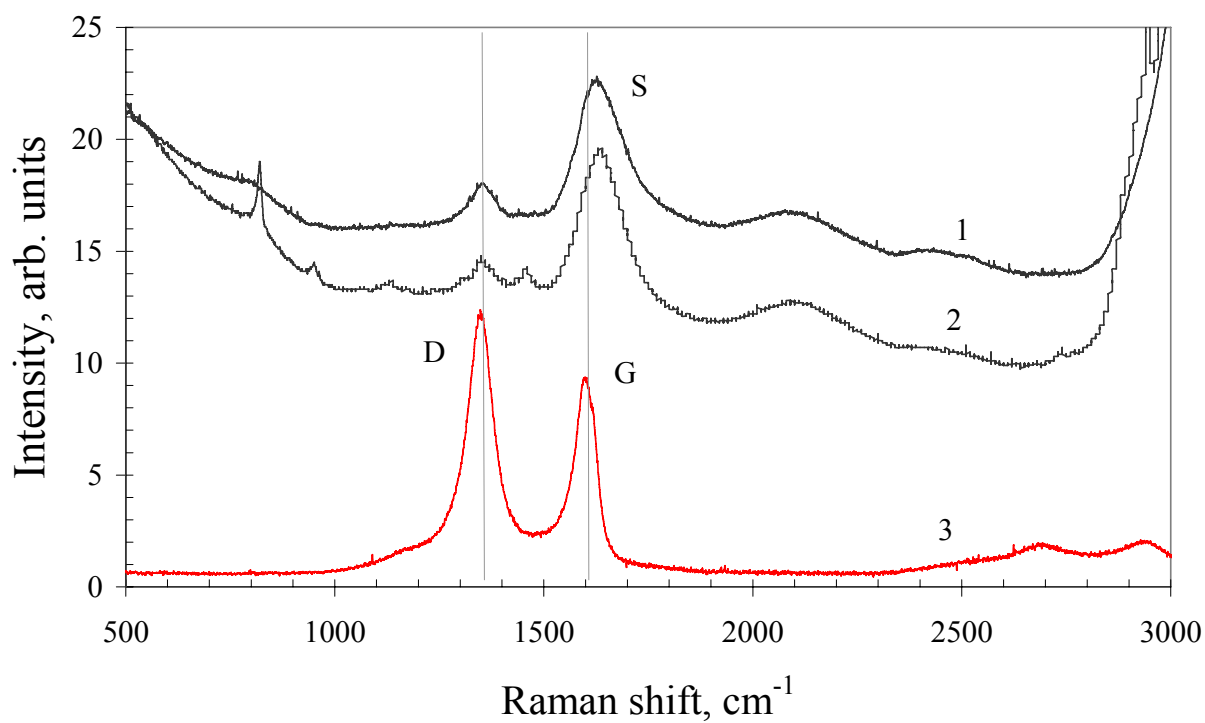

**Figure 2S.** Raman spectra of shungite aqueous dispersions with concentration of 0.06 mg/ml (1) and 0.12 mg/ml (2); and of the dispersion condensate (3). T=293$^0$C, laser excitation at 532 nm.

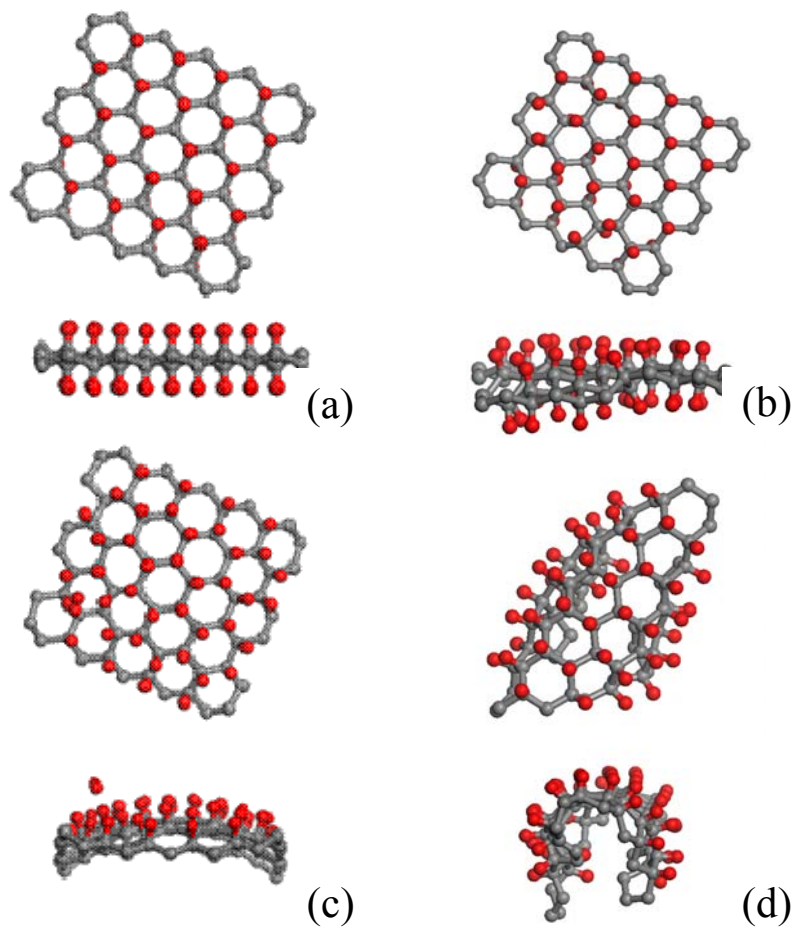

**Figure 3S.** Equilibrium structures (top and side views) of fixed (a, c) and free standing (b, d) (5, 5) NGr membranes under two-side (a, b) and one-side (c, d) hydrogen adsorption. Framing hydrogen atoms are not shown to simplify the structure image presentation. Gray and red balls mark carbon and hydrogen atoms, respectively[6].

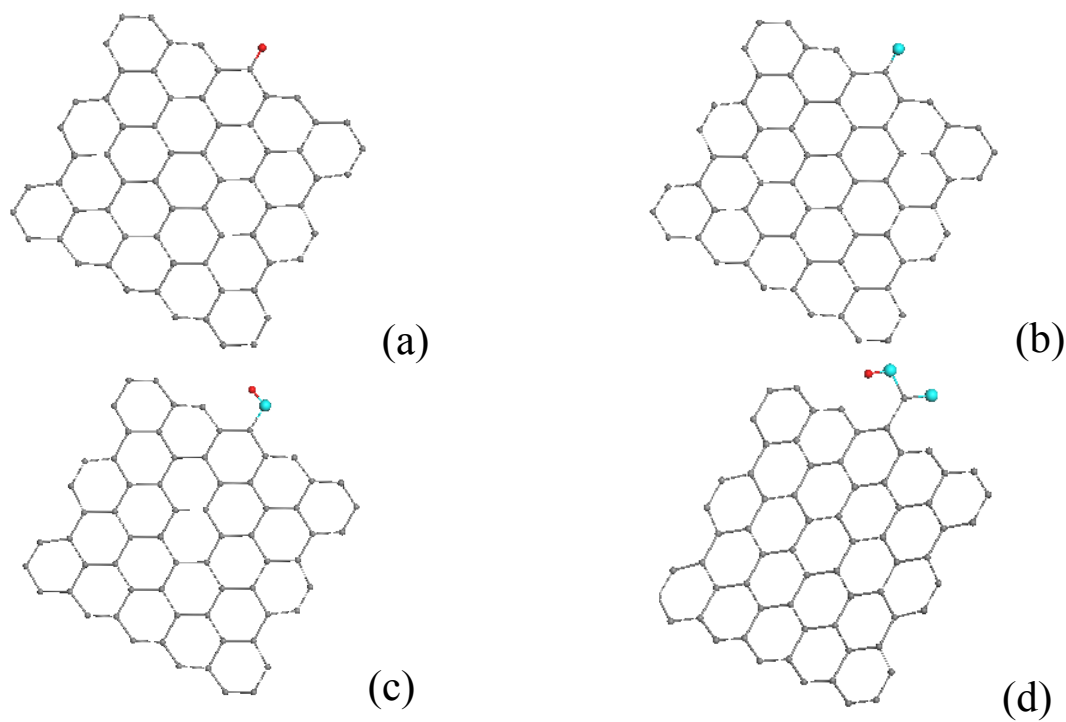

**Figure 4S**. Equilibrium structures of the (5, 5) NGr molecule monohydride (a), O-, OH-, and COOH-monooxides (b, c, and d, respectively)[8].

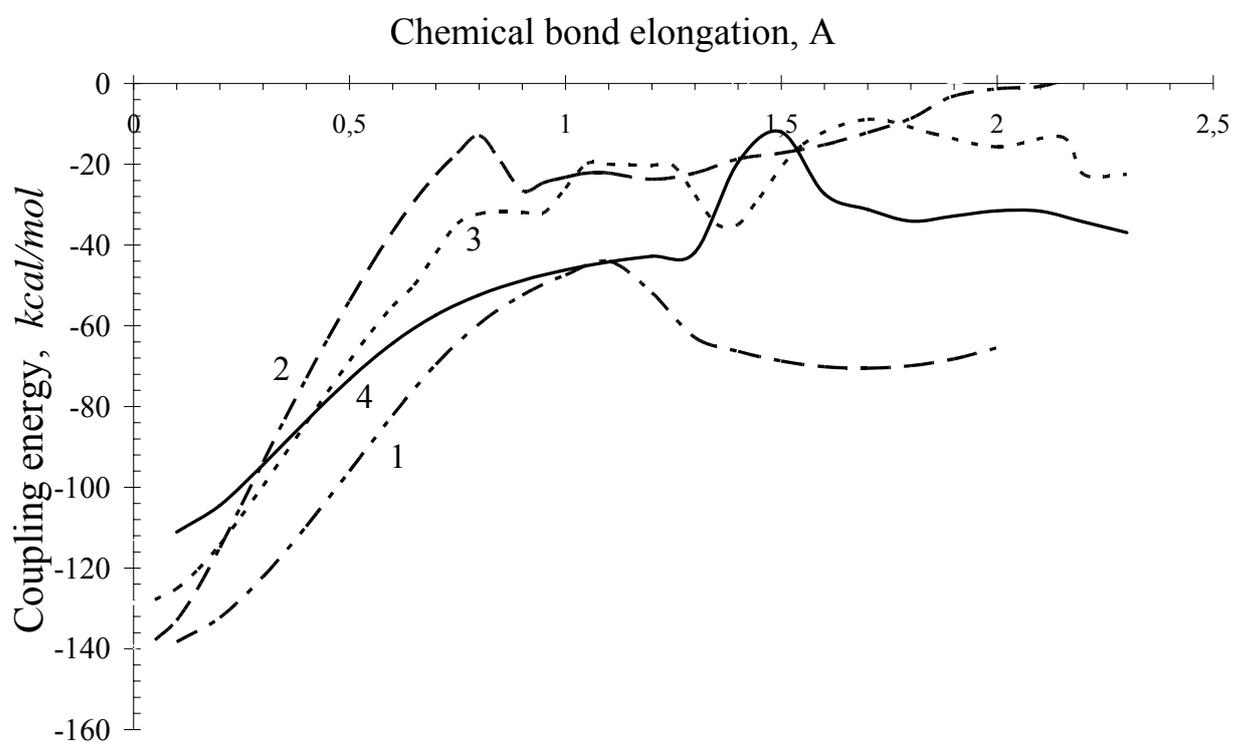

**Figure 5S**. Barrier profiles for desorption of hydrogen (1) and oxygen (2) atoms as well as hydroxyl (3) and carboxyl (4) units from the (5, 5) NGr molecule monohydride and monooxides shown in Fig. 4S.

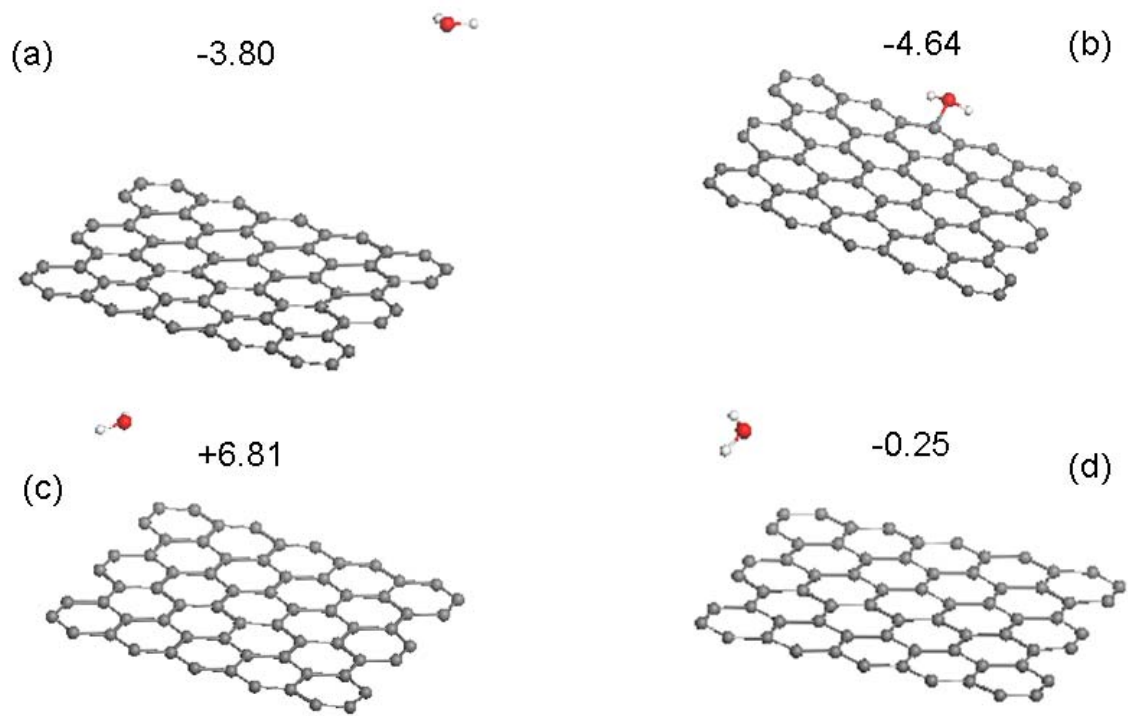

**Figure 6S**. Equilibrium structures of the (5, 5) NGr molecule interacting with one water molecule. Initial positions correspond to C-O distance of 2.02Å (a) and 1.80Å (b) when water is attached to the central zigzag atom; 1.8Å when water is attached to either armchair edge (c) or basal plane (d) atom. Figures indicated the relevant coupling energies in kcal/mol.